\documentstyle[psfig,conf_iap]{article}

\newcommand{\ibid}{\underline{\makebox[0.5in]{}}.}

\newcommand{\kms}{\mbox{km\ s${^{-1}}$}}
\newcommand{\lya}{\mbox{${\rm Ly}\alpha$}}

\begin{document}

\heading{The Galaxy--Absorber Cross-Correlation Function}
 
\par\medskip\noindent

\author{Kenneth M.\ Lanzetta$^1$, John K.\ Webb$^2$, and Xavier Barcons$^3$}
\address{ Department of Physics and Astronomy, State University of New York at
Stony Brook, Stony Brook, NY 11794--3800, U.S.A.}
\address{School of Physics, University of New South Wales, P.O. Box 1,
Kensington, NSW 2033, AUSTRALIA}
\address{Instituto de F{\'\i}sica de Cantabria, Consejo Superior de
Investigaciones Cient{\'\i}ficas, Universidad de Cantabria, Facultad de
Ciencias, 39005 Santander, SPAIN}

\begin{abstract}
We describe an analysis of the galaxy--absorber cross-correlation function
$\xi_{\rm ga}(v,\rho)$ as a function of velocity separation $v$, impact
parameter separation $\rho$, and absorber \lya\ equivalent width $W$ on impact
parameter scales that extend up to $\approx 1 \ h^{-1}$ Mpc.
\end{abstract}

\section{Introduction}

  A primary goal of our imaging and spectroscopic survey of faint galaxies in
fields of Hubble Space Telescope (HST) spectroscopic target QSOs (e.g.\
\cite{L95}) is to determine the gaseous extent of galaxies and the origin of
\lya\ absorption systems by directly comparing redshifts of galaxies and
absorbers identified along common lines of sight.  As the fundamental
statistical measure of the association between galaxies and absorbers, the
galaxy--absorber cross-correlation function bears on this goal for important
practical and theoretical reasons.  From a practical perspective, the
galaxy--absorber cross-correlation function establishes the criterion by which
galaxy--absorber pairs are to be assigned, while from a theoretical perspective,
the galaxy--absorber cross-correlation function constrains the nature of the
relationship between galaxies and absorbers.

  Here we describe an analysis of the galaxy--absorber cross-correlation
function as a function of velocity separation $v$, impact parameter separation
$\rho$, and absorber \lya\ equivalent width $W$.  Full results of the analysis
are described by \cite{L97}.  A dimensionless Hubble constant $h = H_0 / (100 \
{\rm km} \ {\rm s}^{-1} \ {\rm Mpc}^{-1})$ and a deceleration parameter $q_0 =
0.5$ are adopted throughout.

\section{Theoretical Expectations}

  It is easy to show (\cite{L97}) that if there exists a one-to-one
correspondence between at least some galaxies and some absorbers then the
relationship between the galaxy--absorber cross-correlation function and the
galaxy-galaxy auto-correlation function is
\begin{equation}
\xi_{ga}(v,\rho) = \frac{H_0}{m_a} \phi(v,\rho) + f \xi_{gg}(v,\rho),
\end{equation}
where $m_a$ is the absorber number density per unit length, $\phi(v,\rho)$ is
the ``galaxy function'' or the probability of finding an absorber that arises in
a given galaxy within $dv$, $f$ is the fraction of absorbers that arise in
galaxies, and $\xi_{gg}(v,\rho)$ is the galaxy--galaxy autocorrelation function.
Although equation (1) applies {\em only in the absence of absorption line
blending}, it nevertheless serves to illustrate general results.  A more
complicated expression motivated by results of numerical simulations of the
effects of absorption line blending must be applied to determine the actual
relationship between $\xi_{ga}(v,\rho)$ and $\xi_{gg}(v,\rho)$.

  The most important implications of equation (1) are as follows:  (1)  The 
component of $\xi_{ga}(v,\rho)$ related to $\phi(v,\rho)$ exhibits an explicit
dependence on $m_a$.  Because $m_a$ is known to vary as an exponential function
of \lya\ equivalent width $W$, the component of $\xi_{ga}(v,\rho)$ related to
$\phi(v,\rho)$ is expected to vary as an exponential function of $W$.  (2)  At
large impact parameter separations, where $\xi_{gg}(v,\rho) \gg \phi(v,\rho)$,
comparison of $\xi_{ga}(v,\rho)$ and $\xi_{gg}(v,\rho)$ directly yields $f$. 
In practice, however, it is crucial that the effects of absorption line blending
are properly modeled and accounted for in order to determine a meaningful
estimate of $f$.

\section{Observational Material} 

  In summary, the observations consist of (1) optical images and spectroscopy of
objects in the fields of the QSOs, obtained with various telescopes and from the
literature, and (2) ultraviolet spectroscopy of the QSOs, obtained with the HST
using the FOS and accessed through the HST archive.  The optical images and
spectroscopy are used to identify and measure galaxy redshifts and impact
parameters.  The ultraviolet spectroscopy is used to identify and measure
absorber redshifts and equivalent widths.  

  The goal of the present analysis is to investigate galaxies and absorbers that
are cosmologically intervening (rather than associated with the QSOs themselves)
and absorbers that are identified on the basis of \lya\ absorption lines (rather
than on the basis of metal absorption lines).  Hence only galaxies and
\lya-selected absorbers with velocity separations to the QSOs satisfying $v_{\rm
QSO} > 3000$ \kms\ are included into the analysis.  In total, 352 galaxies and
230 absorbers toward 24 fields are included into the analysis, from which 3126
galaxy--absorber pairs are formed.

\section{Analysis}

  The galaxy--absorber cross-correlation function $\xi_{\rm ga}(v,\rho)$ for
all absorbers (i.e.\ with no equivalent width threshold imposed) is shown
versus velocity separation $v$ and impact parameter separation $\rho$ in Figure
1, and the galaxy--absorber cross-correlation function $\xi_{\rm ga}(v,\rho)$
for galaxies at small impact parameters (i.e.\ for impact parameter separations
$\rho < 200 \ h^{-1}$ kpc) is shown versus velocity separation $v$ and
equivalent width threshold $W$ in Figure 2.  Considering Figures 1 and 2
together with similar figures determined over different impact parameter ranges,
with fits of simple models to the observations, and with measurements of the
galaxy-galaxy autocorrelation function $\xi_{gg}(v,\rho)$, the following results
are obtained:

\begin{figure}
\centerline{\vbox{\psfig{figure=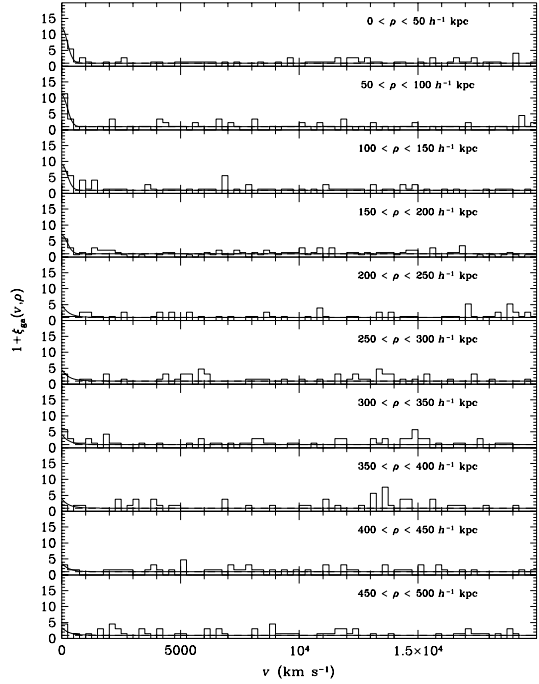,height=10.0cm}}}
\caption[]{Galaxy--absorber cross-correlation function $\xi_{\rm ga}(v,\rho)$
vs.\ velocity separation $v$ and impact parameter separation $\rho$.  Bin size
in velocity separation is 250 \kms, and bin size in impact parameter separation
is $50 \ h^{-1}$ kpc.}
\end{figure}

  (1) Galaxies and absorbers are strongly correlated on velocity scales
$v < 500$ \kms\ and impact parameter scales $\rho < 200 \ h^{-1}$ kpc and are
more weakly correlated on impact parameter scales that range up to at least
$\approx 1 \ h^{-1}$ Mpc.  (2) The amplitude of $\xi_{\rm ga}(v,\rho)$ is
significantly less than the amplitude of the galaxy--galaxy autocorrelation
function on all velocity and impact parameter scales.  (3) The amplitude of
$\xi_{\rm ga}(v,\rho)$ increases roughly exponentially with increasing \lya\
equivalent width threshold on small impact parameter scales (i.e.\ at $\rho <
200 \ h^{-1}$ kpc) but is independent of \lya\ equivalent width threshold (to
within measurement error) on larger impact parameter scales (i.e.\ at $\rho >
200 \ h^{-1}$ kpc).

\begin{figure}
\centerline{\vbox{\psfig{figure=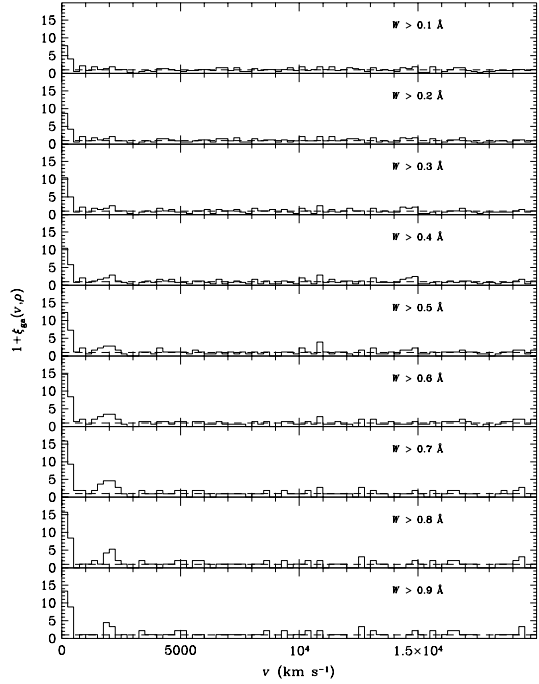,height=10.0cm}}}
\caption[]{Galaxy--absorber cross-correlation function $\xi_{\rm ga}(v,\rho)$
vs.\ velocity separation $v$ and equivalent width threshold $W$ for impact
parameter separations $\rho < 200 \ h^{-1}$ kpc.  Bin size in velocity 
separation is 250 \kms, and step size in equivalent width threshold is $0.1$
\AA.}
\end{figure}

  By combining direct measurements of the galaxy--absorber cross-correlation
function $\xi_{ga}(v,\rho)$  and the galaxy--galaxy auto-correlation function
$\xi_{gg}(v,\rho)$ with equations similar to equation (1) that properly account
for the effects of absorption line blending, it is possible to independently
estimate the galaxy function $\phi(v,\rho)$ and the fraction of absorbers that
arise in galaxies $f$.  The primarily results are that (1) $\phi(v,\rho)$
extends over velocity scales $v < 500$ \kms\ and impact parameter scales $\rho <
200 \ h^{-1}$ kpc and (2) $f$ is compatible with $f = 1$.

\section{Application}

  The most important application of $\xi_{\rm ga}(v,\rho)$ is that it defines a
criterion for statistically distinguishing physical galaxy--absorber pairs from
correlated and random galaxy--absorber pairs.  Physical and correlated pairs are
easily distinguished from random pairs by means of the galaxy--absorber
cross-correlation function $\xi_{\rm ga}(v,\rho)$.  Specifically, a galaxy and
absorber pair of velocity separation $v$ and impact parameter separation $\rho$
is likely to be a correlated or physical pair if the cross-correlation amplitude
satisfies $\xi_{\rm ga}(v,\rho) > 1$ and is likely to be a random pair if the
cross-correlation amplitude satisfies $\xi_{\rm ga}(v,\rho) < 1$.  Physical
pairs are less easily distinguished from correlated pairs, because both can
occur at relatively small velocity and impact parameter separations.  Results
of the previous section indicate the velocity and impact parameter scales over
which $\phi(v,\rho)$ extends, which provides a means of distinguishing physical
pairs from correlated pairs.

  As an application of the practical utility of the galaxy--absorber
cross-correlation function, the relationship between absorber \lya\ equivalent
width $W$ and galaxy impact parameter $\rho$ for galaxy--absorber pairs for
which $\xi_{\rm ga}(v,\rho) > 1$ (i.e.\ for galaxy--absorber pairs that are
likely to be physically associated or correlated rather than random) is shown in
Figure 3.  Figure 3 demonstrates that there exists anti-correlation between
absorber \lya\ equivalent width and galaxy impact parameter, suggesting that the
gas is directly associated with individual galaxies.

\begin{figure}
\centerline{\vbox{\psfig{figure=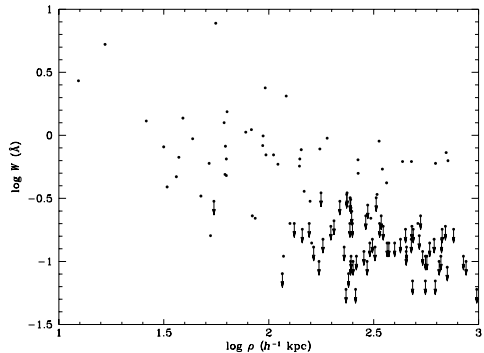,height=5.0cm,angle=-90}}}
\caption[]{Logarithm of \lya\ absorber equivalent width $W$ vs. galaxy impact
parameter $\rho$ for galaxy--absorber pairs for which $\xi_{\rm ga}(v,\rho) > 1$
(i.e.\ for galaxy--absorber pairs that are more likely to be physical or
correlated pairs rather than random pairs).}
\end{figure}

  The anti-correlation between absorber \lya\ equivalent width and galaxy impact
parameter is explored in more detail by \cite{C97a} and \cite{C97b},
considering the possibility of other scaling relationships with galaxy
$B$-band luminosity, morphological type, surface brightness, and redshift.  The
primary result of this analysis is that the gaseous extent of galaxies scales
with galaxy $B$-band luminosity but not with galaxy morphological type, surface
brightness, or redshift.  This indicates that extended gas around galaxies is a
common and generic feature of galaxies spanning a wide range of properties.

\acknowledgements{K.M.L. was supported by NASA grant NAG--53261 and NSF grant
AST--9624216.  X.B. was supported by the DGES under project PB95-00122.}

\vfill
\end{document}